\documentclass[10pt]{iopart}

\usepackage{iopams}
\usepackage{amsfonts,graphicx}

\newcommand{\bra}[1]{\left\langle #1 \right|}
\newcommand{\ket}[1]{\left| #1 \right\rangle}

\newcommand{\rr}{\bi{r}}
\newcommand{\kk}{\bi{k}}
\newcommand{\nn}{\bi{n}}

\newcommand{\reI}{\mathrm{I}}
\newcommand{\reII}{\mathrm{II}}
\newcommand{\reIII}{\mathrm{III}}

\newcommand{\sign}{\mathrm{sign}}

\newcommand{\bopsi}{\psi}
\newcommand{\bosi}{\bsigma}

\begin{document}

\title[Dirac electrons in the presence of matrix potential barrier]{Dirac electrons in the presence of matrix potential barrier: application
to graphene and topological insulators}

\author{Mikhail Erementchouk$^1$, Pinaki Mazumder$^1$, M. A. Khan$^2$ and Michael N. Leuenberger$^2$}
\address{$^1$Department of Electrical Engineering and Computer Science, University of Michigan, Ann Arbor, MI 48109 USA}
\address{$^2$NanoScience Technology Center and Department of Physics, University
of Central Florida, Orlando, FL 32826 USA}
\ead{merement@gmail.com}

\begin{abstract}
Scattering of a 2D Dirac electrons on a rectangular matrix potential barrier is considered using the formalism of spinor transfer matrices. It is shown, in particular, that in the absence of the mass term, the Klein tunneling is not necessarily suppressed but occurs at oblique incidence. The formalism is applied to studying waveguiding modes of the barrier, which are supported by the edge and bulk states. The condition of existence of the uni-directionality property is found. We show that the band of edge states is always finite with massless excitations, while the spectrum of the bulk states, depending on parameters of the barrier, may consist of the infinite or finite band with both, massive and massless, low-energy excitations. The effect of the Zeeman term is considered and the condition of appearance of two distinct energy dependent directions corresponding to the Klein tunneling is found.
\end{abstract}


\section{Introduction}

The unique electron properties of graphene sparked the significant interest in applications of graphene and, more generally, 2D systems supporting massless electron excitations, Dirac electrons. The main challenge in such applications has appeared to stem from the same unique properties. In order to control the electron flow, it is necessary to be able to restrict its motion in desired way and the effect of Klein tunneling (KT) \cite{katsnelson_graphene:_2012,katsnelson_chiral_2006} makes this very difficult: simple scalar potential is not sufficient for keeping the electron from escaping. This circumstance made researchers to consider more general potentials and it was found that Dirac massless electrons can be confined with the help of magnetic barriers \cite{peres_dirac_2006,de_martino_magnetic_2007,tahir_quantum_2008}. This resulted in the significant attention to the problem of propagation of Dirac electrons in the presence of barriers created by both electric and magnetic fields \cite{zhai_theory_2008,ramezani_masir_magnetic_2009,ramezani_masir_kronigpenney_2010,tan_graphene_2010,sharma_electron_2011,ferreira_magnetically_2013,ban_tunable_2015}. Experimentally, the vector potential barrier can be implemented with the help of ferromagnetic stripes with the opposite orientations of magnetization \cite{matulis_wave-vector-dependent_1994,kubrak_magnetoresistance_1999,cerchez_effect_2007} as is illustrated in figure~\ref{fig:general_setup}(a).

Even in the case of piece-wise constant vector and scalar components of the potential, the scattering problem on such barrier turned out to be unexpectedly cumbersome due to the number of parameters characterizing and determining the electron motion and inapplicability of the intuition built by the standard problem of a particle described by the Schrodinger equation scattering on a barrier. As a result, the main analysis of scattering of Dirac electrons on magneto-electric barriers is done for barriers with specific parameters. This makes it difficult to draw the general conclusion about the effect of such barriers on the electron motion.

We consider the problem of the electron propagation in the presence of a rectangular matrix potential barrier and approach it using the developed formalism of spinor transfer matrices. This technique proves to be efficient and allows us to provide the general description of the effect of the barrier. We were able to derive compact expressions relating the reflection and transmission coefficient to the geometry of spinor eigenstates. They show that in the absence of mismatch of the electron mass inside and outside the barrier, the barrier may admit the KT at oblique directions. Moreover, in the case when the barrier is created by the ferromagnetic gate on the surface of a topological insulator, the Zeeman interaction may lead to appearance of two distinct directions corresponding to the KT.

We apply the formalism of spinor transfer matrices to a detailed analysis of waveguiding properties of the matrix potential barrier. This problem considered for the case of combined magnetic-electrostatic barriers on graphene and other 2D materials was the object of consideration of many publications
\cite{ghosh_conductance_2008,masir_wavevector_2008,zhang_guided_2009,myoung_magnetically_2011,huang_graphene_2012,villegas_controlling_2012,ferreira_magnetically_2013,wang_guided_2013,he_guided_2014,he_guided_2015,rickhaus_guiding_2015,xu_guided_2015} with the most attention, however, paid to the bulk states, when the electron states are extended across the barrier. The edge states, with the electron localized near the boundaries of the barrier, appeared only in the context of the mass mismatch \cite{ferreira_magnetically_2013}. Here we show that waveguiding modes based on both, edge and bulk states, can be approached equally. The dispersion equation governing the waveguiding modes can be easily derived using the formalism of the spinor transfer matrices. We analyze the obtained transcendental equations in order to describe the dependence of general properties of the waveguiding modes on parameters of the barrier. In particular, we show that in order to support waveguiding based on edge states the mass mismatch is not required.

The rest of the paper is organized as follows: in Section~\ref{sec:e_states} we introduce spin coherent states in a spatially homogeneous matrix potential, in Section~\ref{sec:tm} we develop the formalism of spinor transfer matrices and apply it for studying scattering on the rectangular barrier, in Section~\ref{sec:waveguiding} we consider waveguiding properties of the barrier, and in Section~\ref{sec:topological} we consider the case when the magnetic field at the boundaries of the barrier affects the electron motion due to the Zeeman term in the Hamiltonian.

\section{Electron states in the presence of the barrier}
\label{sec:e_states}

The equation of motion of the Dirac electron with energy  $\epsilon$ in the presence of rectangular matrix potential barrier $\widehat{U}(x)$ has the
form
\begin{equation}\label{eq:start_eq}
  \left[v {\bi{p}} \cdot \bosi + \widehat{U}(x)\right] \bopsi (\rr) = \epsilon
  \bopsi(\rr),
\end{equation}
where $v$ is the Fermi velocity. In order to shorten formulas, it is
convenient to exclude $v$ by redefining either spatial $r \to r/v$ or energy $\epsilon \to v \epsilon$ scales. Thus, in what follows we take $v = 1$.

\begin{figure}[tb]
  \centering
  \includegraphics[width=3in]{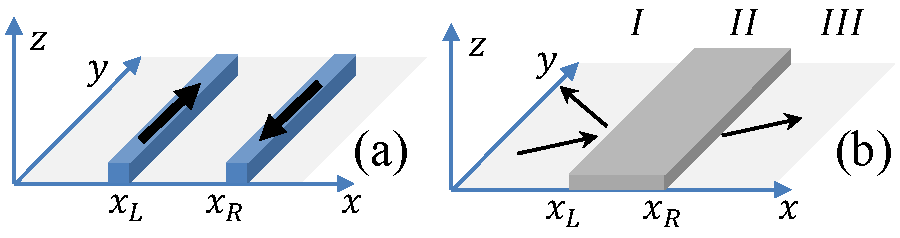}
  \caption{(a) The ferromagnetic gate creating approximately the rectangular barrier $\widehat{U}(x,y) = U_y \theta(x - x_L)\theta(x_R - x) $, where $\theta(x) $ is the Heaviside step function. (b) The schematic depiction of the situation under consideration. The rectangular matrix barrier occupies the region $x_L < x < x_R$, while outside of this region the electron is considered to be free, so that $ \widehat{U}(x,y) =\widehat{U} \theta(x - x_L)\theta(x_R - x) $. }
  \label{fig:general_setup}
\end{figure}

In Eq.~\eref{eq:start_eq} $\bosi = \sigma_x \bi{e}_x + \sigma_y
\bi{e}_y + \sigma_z \bi{e}_z$ is the usual vector of Pauli matrices and
$\bi{p}\cdot \bosi \equiv p_x \sigma_x + p_y \sigma_y$. Employing the fact
that any $2\times 2$ matrix can be expanded over $\{\widehat{1}, \sigma_x,
\sigma_y, \sigma_z\}$, where $\widehat{1}$ is the identity matrix, we
present
\begin{equation}\label{eq:pot_repr}
 \widehat{U}(x) = V(x) \widehat{1} + \bi{U}(x)\cdot \bosi.
\end{equation}
Both $V(x)$ and $\bi{U}(x)$ are assumed to be non-zero only inside the
barrier, $x_L \leq x \leq x_R$, where $V(x) = V$ and $\bi{U}(x) = \bi{U}$,
as illustrated in figure~\ref{fig:general_setup}(b). It should be noted,
however, that arbitrary $U_x(x)$ can be accounted for by the gauge
transformation $\psi(\bi{r}) \to \psi(\bi{r}) \exp\left\{\rmi \int_{x_L}^x dx'
U_x(x')\right\}$. Thus, the effect of this component of the matrix potential
reduces to simple acquiring the phase factor and, therefore, without the
loss of generality one can assume that $U_x(x) \equiv 0$.

The scattering of the Dirac electron on such barrier can be analyzed in the
usual way considering the appropriate solutions in regions $\reI$, $\reII$,
$\reIII$ and imposing the condition of continuity of $\psi(\rr)$ at the
boundaries of the barrier. Within the regions with constant $\bi{U}$ and $V$
the solutions are sought in the form of plane waves $\bopsi(\rr; \kk) =
\bopsi_\kk \rme^{\rmi \kk \cdot \rr}$. The spinor $\bopsi_\kk$ satisfies
\begin{equation}\label{eq:eq_spinor}
  \bosi \cdot \bi{h}(\kk, \bi{U}) \bopsi_\kk = \widetilde{\epsilon}
  \bopsi_\kk,
\end{equation}
where $\widetilde{\epsilon} = \epsilon - V$ and the effective field $\bi{h}$
is defined as
\begin{equation}\label{eq:eff_field}
 \bi{h}(\bi{k},\bi{U}) = \bi{k}  + \bi{U}.
\end{equation}
Equation~\eref{eq:eq_spinor} has the form of the equation for stationary
states of spin $1/2$ in the magnetic field $\bi{h}$. The energies of the
states are
  $\widetilde{\epsilon}_\pm = \pm |\bi{h}|$,
where $|\bi{h}| = \sqrt{h_x^2 + h_y^2 + h_z^2}$. Taking for definiteness
$\epsilon > V$ we obtain
\begin{equation}\label{eq:eps_components}
  \widetilde{\epsilon} = \sqrt{k_x^2  + \left(k_y + U_y \right)^2 + U_z^2}.
\end{equation}

Thus inside regions $\reI$, $\reII$ and $\reIII$ the general solution of
Eq.~\eref{eq:start_eq} is presented as a superposition of $\psi(\rr; \kk)$
corresponding to the same energy $\epsilon$. The invariance with respect to
translations along the $y$-axis implies that $k_y$ is a good quantum number
and therefore scattered states can be characterized by $\epsilon$ and $k_y$. Thus for the given barrier and energy the effective field $\bi{h}$ has definite $|\bi{h}|$, $h_y$ and $h_z$.
On the other hand, the barrier breaks the translational symmetry along
$x$-axis and for given $\epsilon$ and $k_y$ we have two possible values for
$k_x$ corresponding to different signs of $h_x$ keeping $|\bi{h}|$ intact,
\begin{equation}\label{eq:k_x_eq}
  h_x = k_x^{(1,2)} = \pm q,
\end{equation}
where $q^2 = \widetilde{\epsilon}^2 - h_y^2 - h_z^2$. Considering the
scattering of a particle incident on the left boundary of the barrier, the
components with $k_x^{(1)}$ and $k_x^{(2)}$ correspond to the incoming and
reflected state, respectively.

This consideration shows that the representation in terms of the
superposition of plane waves is not trouble free. When $q = 0$ it provides
only one solution, while Eq.~\eref{eq:start_eq} for fixed $k_y$ is
essentially the second order ODE and should have two linearly independent
solutions. Since $q = 0$ is rather an exceptional case we postpone its
detailed discussion to the next section, while for now we assume that $q
\ne0$ and that, indeed, plane wave expansion covers all solutions.

Once the effective fields, $\bi{h}^{(1,2)} = \bi{h}(\kk^{(1,2)}, \bi{U})$,
are found, we can use Eq.~\eref{eq:eq_spinor} to describe the respective
spin states. They are conveniently presented in terms of spin coherent
states \cite{aravind_spin_1999,CombescureCoherent2012}. To vector $\nn$ with Cartesian coordinates
$(\sin(\theta) \cos(\phi), \sin(\theta) \sin (\phi), \cos(\theta))$, where
$\theta$ is the polar angle and $\phi$ is the azimuthal angle, we assign the
state
\begin{equation}\label{eq:spin_coh_def}
  \ket{\nn} = \exp\left[-\rmi \bosi\cdot \bi{m} \theta/2\right] \ket{+},
\end{equation}
where $\bi{m} = (-\sin(\phi), \cos(\phi),0)$ is a unit vector in the
$xy$-plane perpendicular to $\nn$ and $\bi{e}_z$. In terms of amplitudes
with respect to the quantization axis along $\bi{e}_z$ the state $\ket{\nn}$
is
\begin{equation}\label{eq:nn_ampl}
  \ket{\nn} =\left(
               \begin{array}{c}
                 \cos(\theta/2) \\
                 \rme^{\rmi \phi}\sin(\theta/2) \\
               \end{array}
             \right).
\end{equation}
The overlap of two coherent states 
can be presented in a ``covariant" form \cite{aravind_spin_1999}
\begin{equation}\label{eq:coh_overlap}
  \langle \nn | \nn' \rangle = \left[\frac{1}{2} (1 + \nn\cdot
  \nn')\right]^{1/2}
  \exp\left[\frac{i}{2} A(\nn, \nn')\right],
\end{equation}
where $A(\nn, \nn') = A(\nn, \nn', \bi{e}_z)$ is the oriented area of the
spherical triangle with vertices at $\nn$, $\nn'$ and $\bi{e}_z$.

When all components of $\bi{h}$ are real numbers (i.e. when $q^2 >0$),
solutions of Eq.~\eref{eq:eq_spinor} have the simple form:
$\ket{\nn^{(1,2)}}$ with $\nn^{(1,2)} =
\bi{h}^{(1,2)}/|\widetilde{\epsilon}|$. It should be noted that the condition $q^2 > 0$ holds only when $|\widetilde{\epsilon}| > 0 $, thus the direction of the spin coherent state in this case is always well-defined.

The situation is more complex when $q^2 < 0$. In this case $\bi{h}$ has imaginary $x$-component and the orientation of the coherent state should be found directly from
Eq.~\eref{eq:eq_spinor}. First, we assume that $h_y >0$ and $U_z = 0$ and
then extend the consideration to the general case. Let $h_x = i \kappa$,
then Eq.~\eref{eq:eq_spinor} can be written as
\begin{equation}\label{eq:spinor_eq_comp}
\eqalign{
  - \widetilde{\epsilon} \psi_1 - \rmi (h_y - \kappa) \psi_2 = 0, \cr
  \rmi (h_y + \kappa) \psi_1 - \widetilde{\epsilon} \psi_2 = 0.}
\end{equation}
Taking into account that $\widetilde{\epsilon}^2 = h_y^2 - \kappa^2$ (notice
that $\kappa^2 < h_y^2)$ we find
\begin{equation}\label{eq:components_decay}
  \frac{\psi_2}{\psi_1} = \rmi \, \sign(\widetilde{\epsilon}) \sqrt{\frac{h_y +
  \kappa}{h_y - \kappa}}.
\end{equation}
Comparing with Eq.~\eref{eq:nn_ampl} we can see that
Eq.~\eref{eq:components_decay} describes a state $\ket{\nn}$ with $\nn$
lying in the $yz$-plane. The polar angle of $\nn$ can be presented as
$\theta = \pi/2 + \Delta \theta$, where $\Delta\theta$ is the angle of
deviation from the $y$-axis and is found to be
\begin{equation}\label{eq:swing_att}
  \tan(\Delta\theta) = \frac{\kappa}{\widetilde{\epsilon}}.
\end{equation}
The azimuthal angle is $\phi = \pi/2$ when $\epsilon > V$ and $\phi =
-\pi/2$ otherwise. The second solution, corresponding to $h_x = - i \kappa$
is found by simple reversing the sign of $\kappa$. Thus, it is characterized
by the same azimuthal angle but is deviated from the $y$-axis down.

The case with arbitrary sign of $h_y$ and $U_z \ne 0$ can be studied using
the same approach. For this we rotate the coordinate system for
Eq.~\eref{eq:eq_spinor} around the $x$-axis in such way that the transformed
$y$-axis is oriented along the projection of $\bi{h}$ on the $yz$-plane,
that is along the vector $\bi{h}_{yz} =(0, h_y, h_z)$. In these coordinates
Eq.~\eref{eq:eq_spinor} takes the same form as \eref{eq:spinor_eq_comp} with
$\sqrt{h_y^2 + h_z^2}$ instead of $h_y$. The polar angle of $\nn$ is
presented then as $\theta = \theta_0 + \Delta \theta$, where $\theta_0$ is
the polar angle of $\bi{h}_{yz}$ and $\Delta \theta$ is determined by
Eq.~\eref{eq:swing_att}. The azimuthal angle, in turn, depends on the sign
of $h_y$: if $h_y > 0$, then $\phi$ is determined by the same rule as above:
$\phi = \sign(\widetilde{\epsilon})\pi/2$; if, however, $h_y < 0$, then the
rule is reversed $\phi = -\sign(\widetilde{\epsilon})\pi/2$.

\begin{figure}
  \centering
  \includegraphics[width=2in]{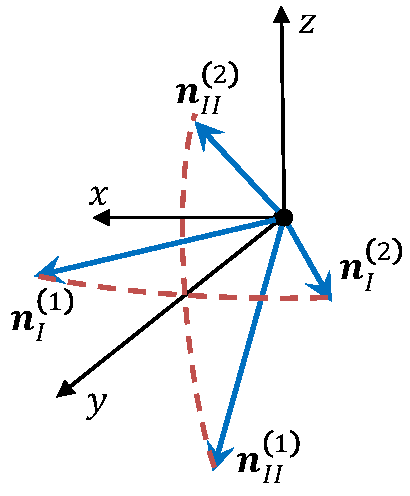}\\
  \caption{Directions of spins corresponding to $h_x^2 > 0$
  (spins are in the $xy$-plane) and $h_x^2 < 0$ (spins are
  in the $yz$-plane) for the case $U_z=0$. When $U_z\ne0$ one should
  substitute
  $\bi{h}_{yz}$, the projection of the effective field on the $yz$-plane,
  in place of the $y$-axis.}\label{fig:spinors}
\end{figure}

Thus, roughly speaking, when $h_x$ is imaginary its magnitude determines the
deviation of the spinor from the direction of vector $\bi{h}_{yz}$ in the
$yz$-plane (see figure~\ref{fig:spinors}).

As we can see, when $q^2 < 0$, for a given energy $\epsilon$ we as well have
two states  characterized by $k_x^{(1,2)} = \pm \rmi |q|$ and spins oriented
along $\nn^{(1,2)}$. We enumerate the solutions in such way that $k_x^{(1)}$
corresponds to exponentially decaying state with increasing the penetration
depth, while $k_x^{(2)}$ corresponds to the exponentially growing one.

An important symmetry of vectors $\nn^{(1)}$ and $\nn^{(2)}$ representing
spinor states should be noted. For both cases, $q^2 > 0$ and $q^2 < 0$,
vectors $\nn^{(1)}$, $\nn^{(2)}$ and $\bi{h}_{yz}$ lie in the same plane,
and $\nn^{(1)}$ and $\nn^{(2)}$ are related through reflection about
$\bi{h}_{yz}$  in this plane. This symmetry will be extensively used below.

In order to formally manifest the symmetry it is convenient to present
eigenstates of Eq.~\eref{eq:eq_spinor} using dilation operators. For the
case $q^2 > 0$, we have
\begin{equation}\label{eq:n_dil_prop}
  \ket{\nn^{(1,2)}} = K^{(1,2)} \exp\left(b \bi{h}_{yz} \cdot \bosi /2 \right)
  \ket{\pm \bi{e}_x},
\end{equation}
where $K^{(1,2)} = \rme^{\pm \rmi \phi/2}\sqrt{q/|\widetilde{\epsilon}|}$,
$\cosh(b h_{yz})=|\widetilde{\epsilon}|/q$, $\sinh(b h_{yz}) =
\sign(\widetilde{\epsilon}) h_{yz}/q$. When $q^2 < 0$, so that $q = \rmi
\kappa$ with $\kappa > 0$, we obtain
\begin{equation}\label{eq:n_dil_att}
 \ket{\nn^{(1,2)}} = C^{(1,2)} \exp\left(b \bi{h}_{yz} \cdot \bosi /2 \right)
 \ket{\pm \bi{l}},
\end{equation}
where $\bi{l} = \bi{h}_{yz} \times \bi{e}_x/h_{yz}$, $C^{(1)} =
\sqrt{\kappa/h_{yz}}$, $C^{(2)} = \rmi \sqrt{\kappa/h_{yz}}$, $\cosh(b
h_{yz}) = h_{yz}/\kappa$, $\sinh(b h_{yz}) = \widetilde{\epsilon}/\kappa$.

\section{Transfer matrix approach for spinors}
\label{sec:tm}

The analysis above shows that Eq.~\eref{eq:eq_spinor} can be regarded as
defining two distributions of directions $\nn^{(1)}(x)$ and $\nn^{(1)}(x)$,
corresponding to forward and backward propagating
modes. The spatial inhomogeneity of $\bi{U}(x)$ and $V(x)$ together with the
continuity condition couple these distributions leading to scattering, which
is conveniently described by the formalism of transfer-matrices.

Inside regions $\reI$, $\reII$ and $\reIII$ we have
\begin{equation}\label{eq:solutions_inside}
  \psi_i(x) = \sum_{j=1}^2 \widetilde{\alpha}_i^{(j)} \ket{\nn^{(j)}_{i}} 
  \rme^{\rmi  k^{(j)}_{i;x} x}
\end{equation}
where $i$ runs over $\{\reI,\reII,\reIII\}$, $\widetilde{\alpha}_i^{(j)}$
are some complex amplitudes and we have omitted the common factor $\exp(i
k_y y)$.

First, let us consider two points $x_a$ and $x_b$ arranged as $x_a \leq x_b
< x_L$. One can see that $\psi(x_b)$ differs from $\psi(x_a)$ by phase
factors acquired by amplitudes $\widetilde{\alpha}_\reI^{(1,2)}$. We present
the relation in the form
\begin{equation}\label{eq:inside_tm_eq}
  \left(
    \begin{array}{c}
      \widetilde{\alpha}_\reI^{(1)} \\
      \widetilde{\alpha}_\reI^{(2)} \\
    \end{array}
  \right)_{x_b} = \widehat{T}_{\reI,\reI}(x_b - x_a)
  \left(
    \begin{array}{c}
      \widetilde{\alpha}_\reI^{(1)} \\
      \widetilde{\alpha}_\reI^{(2)} \\
    \end{array}
  \right)_{x_a},
\end{equation}
which defines the transfer matrix within region $\reI$
\begin{equation}\label{eq:outside_tm}
  \widehat{T}_{\reI,\reI}(x) =
  \left(
    \begin{array}{cc}
      \rme^{\rmi k^{(1)}_{\reI;x}x} & 0 \\
      0 & \rme^{\rmi k^{(2)}_{\reI;x}x} \\
    \end{array}
  \right).
\end{equation}
In the similar way the transfer matrix $\widehat{T}_{\reII,\reII}(x)$ within
region $\reII$ can be defined. It has the same form as
$\widehat{T}_{\reI,\reI}(x)$ but with $k_{\reI;x}^{(1,2)}$ replaced by
$k_{\reII;x}^{(1,2)}$. In order to simplify notations we denote $k_{\reII;x}^{(1,2)} = \pm q$ with $q^2 = \widetilde{\epsilon}^2 -
h_{\reII; y}^2 - h_{\reII; z}^2$. Thus, the transfer matrix within the
barrier has the form
\begin{equation}\label{eq:inside_tm}
  \widehat{T}_{\reII, \reII}(x) =
  \left(
    \begin{array}{cc}
      \rme^{\rmi q x} & 0 \\
      0 & \rme^{-\rmi q x} \\
    \end{array}
  \right).
\end{equation}

The form of the transfer matrices inside the regions allows us to
incorporate phases at the boundaries of the barrier into the amplitudes and
define $\alpha_{\reI}^{(j)} = \widetilde{\alpha}_{\reI}^{(j)} e^{i
k_{\reI;x}^{(j)} x_1}$, $\alpha_{\reII}^{(j)} =
\widetilde{\alpha}_{\reII}^{(j)} e^{i k_{\reII;x}^{(j)} x_1}$ and
$\alpha_{\reIII}^{(j)} = \widetilde{\alpha}_{\reIII}^{(j)} e^{i
k_{\reIII;x}^{(j)} x_2}$. In other words, except for $\alpha_{I}^{(j)}$ we
have included into amplitudes their phases at the outmost left points of
discontinuity of the potential.

In terms of such amplitudes the continuity condition at the left boundary of
the barrier takes a simple form
\begin{equation}\label{eq:tbe_eq}
  \alpha_{\reI}^{(1)} \ket{\nn_{\reI}^{(1)}} + \alpha_{\reI}^{(2)}
  \ket{\nn_{\reI}^{(2)}} =
  \alpha_{\reII}^{(1)} \ket{\nn_{\reII}^{(1)}} + \alpha_{\reII}^{(2)}
  \ket{\nn_{\reII}^{(2)}}
\end{equation}
and can be presented as
\begin{equation}\label{eq:tbe_def}
  \left(
    \begin{array}{c}
      \alpha_{\reII}^{(1)} \\
      \alpha_{\reII}^{(2)} \\
    \end{array}
  \right) = \widehat{T}_{\reII,\reI}
    \left(
    \begin{array}{c}
      \alpha_{\reI}^{(1)} \\
      \alpha_{\reI}^{(2)} \\
    \end{array}
  \right),
\end{equation}
where $\widehat{T}_{\reII,\reI}$ is the transfer matrix through the
interface between the free space and the barrier. This shows the distinctive feature of the formalism of spinor transfer matrices compared with usually employed transfer matrices for amplitudes of the waves propagating to the left and to the right. The latter relates the spinor amplitudes in the chosen basis, which hides the structure of the eigenstates under the relation between up- and down-components in the chosen basis, while, of course, formally representing the same electron wavefunction. The spinor transfer matrices, in turn, relate the amplitudes of the local eigenstates thus describing the propagation of the electron in ``covariant'' terms. As will be demonstrated below, this simplifies significantly the analysis of the scattering on the barrier.

If $\det(\widehat{T}_{\reII,\reI}) \ne 0$, that is
$\ket{\nn_{\reII}^{(1)}}$ and $\ket{\nn_{\reII}^{(2)}}$ do not coincide
(the meaning of this condition will be discussed in details below), one can
easily check the relation
\begin{equation}\label{eq:ttot_def}
\left(
    \begin{array}{c}
      \alpha_{\reIII}^{(1)} \\
      \alpha_{\reIII}^{(2)} \\
    \end{array}
  \right) = \widehat{T}_{\mathrm{tot}}
    \left(
    \begin{array}{c}
      \alpha_{\reI}^{(1)} \\
      \alpha_{\reI}^{(2)} \\
    \end{array}
  \right),
\end{equation}
with $\widehat{T}_{\mathrm{tot}} = \widehat{T}_{\mathrm{\reII,\reI}}^{-1}
\widehat{T}_{\reII,\reII}(d) \widehat{T}_{\reII,\reI}$, where $d = x_2 -
x_1$ is the width of the barrier.

This consideration can be generalized straightforwardly to the case of
multiple barriers: for each interface between regions with constant
potential and magnetic field one finds the respective transfer matrix from
an equation similar to Eq.~\eref{eq:tbe_eq}, while propagation inside the
regions is described by diagonal matrices similar to
$\widehat{T}_{\reII,\reII}$. It should be noted that matrix
$\widehat{T}_{\reII,\reII}$ takes the same form also in the case with
exponentially decaying and growing solutions.

Finally, once the total transfer matrix is known one can find the reflection
and transmission amplitudes for incidence from the left by solving the
equation $(t, 0)^T = \widehat{T}_{\mathrm{tot}} (1, r)^T$ and for incidence
from the right from $(r', 1)^T = \widehat{T}_{\mathrm{tot}} (0,t')^T$. The
structure of the total transfer matrix imposes some general limitations on
the reflection and transmission amplitudes. In particular, it can be shown
that the reflection and transmission amplitudes in the direct and reverse
directions may differ at most by a phase factor.

Now we turn to solving Eq.~\eref{eq:tbe_eq} and finding the transfer matrix
through the boundary of the barrier. We would like to notice that
Eq.~\eref{eq:tbe_eq} has the form of presenting the same spinor in bases
defined by pairs $\ket{\nn_{\reI}^{(1,2)}}$ and $\ket{\nn_{\reII}^{(1,2)}}$.
Thus, $\widehat{T}_{\reII,\reI}$ has the meaning of a matrix describing the
transformation between different, not necessarily orthogonal, bases. 
The transfer matrix is found by employing the dual basis. We define
$\bra{\overline{\nn_i^{(j)}}}$ in such way that
$\bra{\overline{\nn_i^{(j)}}} \left. \nn_i^{(l)} \right \rangle =
\delta_{jl}$. Thus $\bra{\overline{\nn_i^{(1)}}} =
       \bra{-\nn_i^{(2)}}\left. \nn_i^{(1)} \right \rangle^{-1}
       \bra{-\nn_i^{(2)}}$ and $\bra{\overline{\nn_i^{(2)}}} =
       \bra{-\nn_i^{(1)}}\left. \nn_i^{(2)} \right \rangle^{-1}
       \bra{-\nn_i^{(1)}}$.
Using these definitions the interface transfer matrix is found to be
\begin{equation}\label{eq:tbe_form}
 \widehat{T}_{\reII,\reI} =
 \left(
   \begin{array}{cc}
     \bra{\overline{\nn_{\reII}^{(1)}}} \left. \nn_{\reI}^{(1)} \right \rangle
     &
            \bra{\overline{\nn_{\reII}^{(1)}}} \left. \nn_{\reII}^{(2)} \right
            \rangle \\
     \bra{\overline{\nn_{\reII}^{(2)}}} \left. \nn_{\reI}^{(1)} \right \rangle
     &
            \bra{\overline{\nn_{\reII}^{(2)}}} \left. \nn_{\reI}^{(2)} \right
            \rangle \\
   \end{array}
 \right).
\end{equation}

Due to the mutual arrangement of $\nn_i^{(j)}$ the form of
$\widehat{T}_{\mathrm{II,I}}$ is far from arbitrary. When $q^2
> 0$, we have
\begin{equation}\label{eq:tbe_form1}
   \widehat{T}_{\reII,\reI} =
  \left(
    \begin{array}{cc}
      a \rme^{\rmi \alpha} &
         b \rme^{-\rmi \beta} \\
      b \rme^{\rmi \beta} & a \rme^{-\rmi \alpha}\\
    \end{array}
  \right),
\end{equation}
and when $q^2 <0$
\begin{equation}\label{eq:tbe_form2}
   \widehat{T}_{\reII,\reI} =
  \left(
    \begin{array}{cc}
      a \rme^{\rmi \alpha} &
         a \rme^{-\rmi \alpha} \\
      b \rme^{\rmi \beta} & b \rme^{-\rmi \beta}\\
    \end{array}
  \right),
\end{equation}
where
\begin{equation}\label{eq:coeff_t}
\eqalign{
 a = \left | \left \langle \overline{\nn_{\reII}^{(1)}} \right.
 \ket{\nn_{\reI}^{(1)}} \right| =
    \sqrt{\frac{1-\nn_{\reII}^{(2)} \cdot \nn_{\reI}^{(1)}}{1-\nn_{\reII}^{(2)}
    \cdot \nn_{\reII}^{(1)}}}, \cr
 b = \left | \left \langle \overline{\nn_{\reII}^{(2)}} \right.
 \ket{\nn_{\reI}^{(1)}} \right| =
    \sqrt{\frac{1-\nn_{\reII}^{(1)} \cdot \nn_{\reI}^{(1)}}{1-\nn_{\reII}^{(2)}
    \cdot \nn_{\reII}^{(1)}}}
}
\end{equation}
and
\begin{equation}\label{eq:coeff_t_ph}
\eqalign{
 \alpha = \frac{1}{2} A(-\nn_{\reII}^{(2)},\nn_{\reI}^{(1)}) - \frac{1}{2}
 A(-\nn_{\reII}^{(2)}, \nn_{\reII}^{(1)}), \cr
 \beta = \frac{1}{2} A( -\nn_{\reII}^{(1)},\nn_{\reI}^{(1)}) + \frac{1}{2}
 A(-\nn_{\reII}^{(2)}, \nn_{\reII}^{(1)}).
}
\end{equation}

When there are no
propagating modes either inside and outside of the barrier, i.e. when
$\left(k_{\reI;x}^{(1,2)}\right)^2 < 0$ and $q^2 < 0$,  
$\widehat{T}_{\reII,\reI}$ has form \eref{eq:tbe_form2} with $\alpha = \beta
= 0$.

Taking into account the general form of the transfer matrices we find for
the case $q^2 >0$
\begin{equation}\label{eq:transp_case1}
\eqalign{
 r & = \frac{2i}{D} \rme^{\rmi(\alpha + \beta)}
    ab \sin( qd), \cr
 t & = \frac{1}{D} (a^2-b^2),}
\end{equation}
where $D = a^2 \rme^{-\rmi qd} - b^2 \rme^{\rmi qd}$. In order to analyze the reflection
and transmission properties closer it is convenient to employ the general
property $|r|^2 + |t|^2 = 1$ and to consider
\begin{equation}\label{eq:rt_rat}
\eqalign{
  \left | \frac{r}{t}\right|^2  = &
    \frac{\left(1-\nn_{\reII}^{(2)} \cdot \nn_{\reI}^{(1)} \right)
            \left(1-\nn_{\reII}^{(1)} \cdot \nn_{\reI}^{(1)} \right)}{
            \left[ \nn_{\reI}^{(2)} \cdot \left(\nn_{\reII}^{(1)} -
            \nn_{\reII}^{(2)} \right) \right]^2}
        \sin^2(qd) \cr
& = \frac{(\epsilon U_y + V k_y )^2 + U_z^2 k_{\reI; x}^2}{4 k_{\reI; x}^2 q^2}
\sin^2(qd) .
}
\end{equation}
Here we have taken into account that $\nn_{\reII}^{(1,2)} \cdot
\nn_{\reI}^{(1)} = \bi{h}_{\reII}^{(1,2)} \cdot \bi{h}_{\reI}^{(1)}/\epsilon
\widetilde{\epsilon}$ and, therefore, Eq.~\eref{eq:rt_rat} is valid for an
arbitrary relation between $\epsilon$ and $V$.

Equations~\eref{eq:transp_case1} and \eref{eq:rt_rat} clearly distinguish
between the effects of mismatch of directions of the effective fields inside
and outside the barrier and the effect of interference due to scattering
from the front and back sides of the barrier. In particular, one can see
that the reflection coefficient vanishes when either
\begin{equation}\label{eq:}
  \sin\left(q d \right) = 0,
\end{equation}
or when
\begin{equation}\label{eq:KT_cond}
  \nn_{II}^{(1)} \cdot \nn_{I}^{(1)} = 1.
\end{equation}
The first condition is responsible for the periodic variation of the reflection
coefficient with the width of the barrier due to the interference effect. The second condition
is satisfied when directions of the spins inside and outside the barrier
coincide. In this case the reflection coefficient is zero regardless the width
of the barrier and thus is associated with the KT.

\begin{figure}[tb]
  \centering
  \includegraphics[width=4 in]{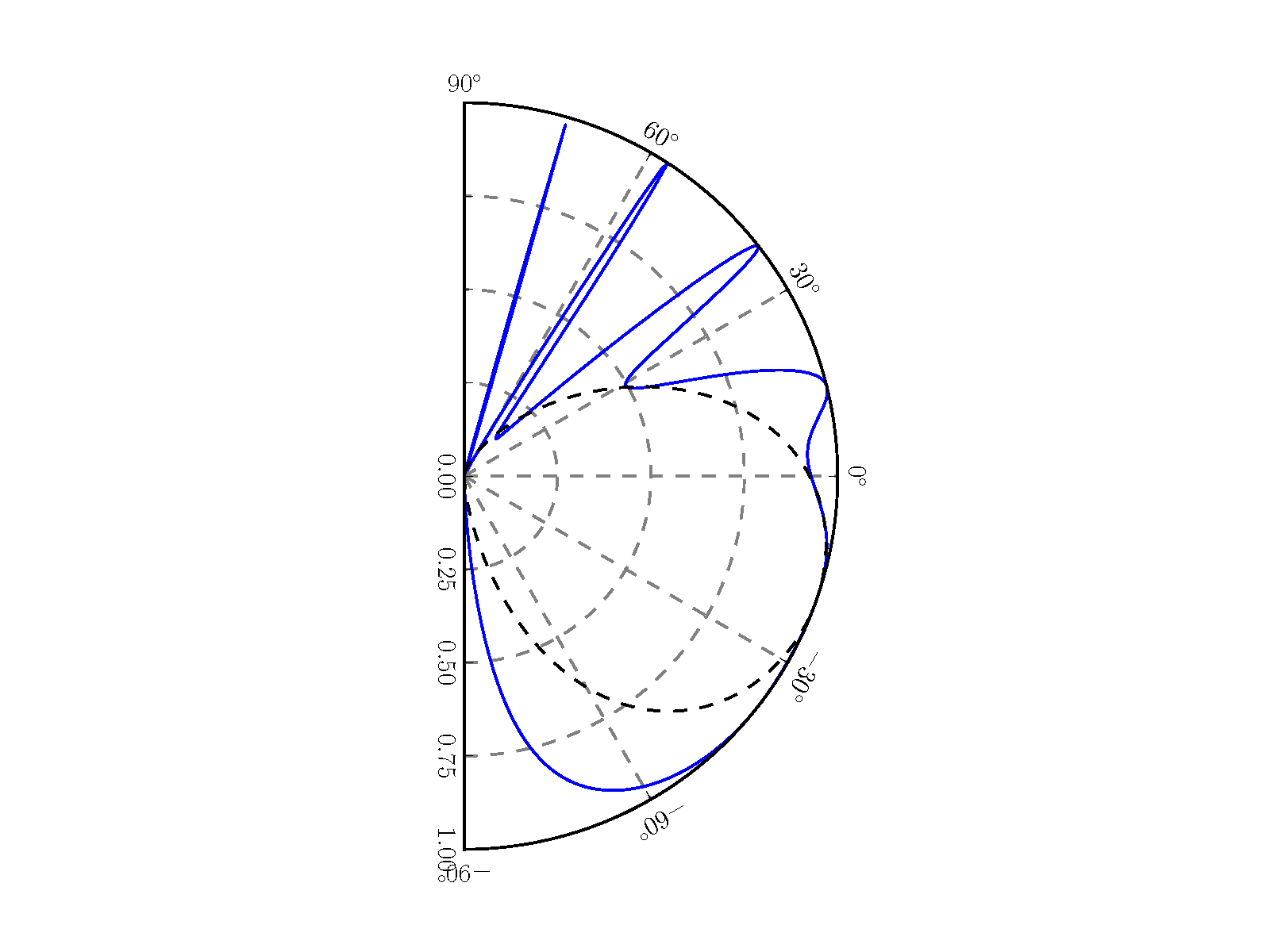}
  \caption{The angular dependence of the transmission coefficient through the barrier with $U_y/\epsilon = 1$, $V/\epsilon = 3$ for two different widths: (dashed line) $\epsilon d = 1$ and (solid line) $\epsilon d = 10$ in the units adopted in the main text.  For the magnetic barrier with $B = 0.5$ T the spectra correspond to $\epsilon = 19$ meV, $V = 57$ meV and $d = 37$ nm and $d = 370$ nm, respectively. The KT takes place at oblique incidence with $\theta = \arcsin(-V/U_y) \approx -20^\circ $.
  }
  \label{fig:angular}
\end{figure}

Obviously, condition \eref{eq:KT_cond} cannot be satisfied when $U_z \ne 0$.
Thus the respective barriers (often called mass barriers) completely suppress the KT. The effect of $V$
and $U_y$ on the KT is less straightforward. It follows from
Eq.~\eref{eq:rt_rat} that in the case $U_z = 0$ the KT takes place when
\begin{equation}\label{eq:KT_cond_obl}
  k_y = - \frac{U_y \epsilon}{V}.
\end{equation}
Thus, when the barrier contains both $V$ and $U_y$, the KT is not
necessarily suppressed but may appear for obliquely incident Dirac electron as is illustrated in figure~\ref{fig:angular}.

With increasing $U_y$ the Klein direction deviates more from normal until it
becomes parallel to the boundary of the barrier. Further increase of $U_y$
will lead to suppression of the KT. Thus, in order for KT to exist, $U_y$
and $V$ in the barrier must satisfy
\begin{equation}\label{eq:KT_cond_fields}
  |U_y| < |V|.
\end{equation}

We would like to remark that if the KT condition is satisfied for an
electron incident with $k_y \ne0$ from the left, then the condition is not
fulfilled for the electron with time reversed trajectory. In the latter case
$k_y$ has the opposite sign and Eq.~\eref{eq:KT_cond_obl} no longer holds.
In order to recover the KT the full time reversal transformation must be performed, 
which includes reversing $\bi{U}$.

This analysis directly applies to an electron in graphene in the presence of
scalar and vector potential barrier. In this case $V(x)$ has the meaning of the
scalar potential and $U_x$ and $U_y$ are the respective components of the
vector potential, which creates the magnetic field $\bi{B}(x) = -B l_B \bi{e}_z
[\delta(x - x_L) - \delta(x - x_R)]$, where $l_B = \sqrt{\hbar/e B}$, $B_0 l_B = U_y/e$ and $e$ is the magnitude of the electron charge. The effect of such magnetic barrier on the KT was studied in Refs.~\cite{zhai_theory_2008,sharma_electron_2011}. In \cite{zhai_theory_2008} the KT at oblique directions was observed numerically, while in \cite{sharma_electron_2011} it was concluded that the addition of the magnetic barrier to the scalar potential barrier suppresses the KT. Our consideration above resolves unambigously this controversy. The magnetic barrier alone, indeed, doesn't demonstrate the KT. However, when it is accompanied with the scalar potential such that condition~\eref{eq:KT_cond_fields} is fulfilled, the KT restores
at oblique incidence. 

It should be emphasized that the KT takes place when the direction of the eigen-spinors is uniform across the system. In the non-attenuated regime, i.e. when $q^2 >0$, this is equivalent to a uniform distribution of the directions of the effective field, $\bi{h}(x)/\widetilde{\epsilon}(x) = \bi{n}(x) = \mathrm{const}$. Thus the oblique KT is the local property of the matrix potential governing the motion of the Dirac electron and, therefore, it holds for barriers with more
complex spatial variation of the potentials. Evidently, if $U_y(x)/V(x) = c < 1$ is a constant across the
barrier, then such barrier is reflection-less for electrons incident at the
angle $\chi = -\arcsin(c)$ counted from the normal to the boundary. Conversely, this shows that in the barriers with a general spatial variation of the scalar and vector components the KT condition is, generally speaking, broken. For example, in anti-parallel ferromagnetic gates of finite width, $\Delta x$, the vector potential varies continuously and, except when the scalar potential is carefully chosen to satisfy $U_y(x)/V(x) = c < 1$, the condition $\bi{n}(x) = \mathrm{const} $ doesn't hold. 

A detailed analysis of a general coordinate dependent matrix potential goes beyond the scope of the present paper. We limit ourselves to a qualitative discussion of the case of thin gates, such that $ \mathrm{max}(\epsilon, \widetilde{\epsilon}) \cdot \Delta x \ll 1 $. The effect of the gradual variation of the matrix potential at, say, the left boundary of the barrier is taken into account by replacing $\widehat{T}_{\reII,\reI}$ in the expression for $\widehat{T}_{\mathrm{tot}}$ by $\widehat{T}_{\reII,\reI} \widehat{T}_{\mathrm{i}}$, where $\widehat{T}_{\mathrm{i}}$ is the transfer matrix through the barrier described by $\Delta \widehat{U}(x) = \widehat{U}(x) - \widetilde{U}(x)$, where $\widetilde{U}(x)$ is the full potential and $\widehat{U}(x)$ is its approximation by the rectangular barrier near its left boundary. Thus, $\Delta \widehat{U}(x)$ differs from zero only inside the gate. Then it can be seen that along the direction corresponding to the KT for the rectangular barrier, the reflection coefficient doesn't vanish identically but is an oscillating function of the width of the barrier $|r|^2 = 4|r_{\mathrm{i}}|^2 \sin^2 (qd) $, where $r_{\mathrm{i}}$ is the reflection coefficient of the single barrier described by $\Delta \widehat{U}(x)$ or, equivalently, of the single gate at the KT direction. Here we have taken $\widehat{T}_{\mathrm{i}}$ in the first nonvanishing order of $\Delta x$
\begin{equation}\label{eq:transition_TM}
 \widehat{T}_{\mathrm{i}} = \widehat{1} + i \int dx \widehat{Q}(x),
\end{equation}
where $\left[ \widehat{Q}(x) \right]_{j,l}  = \bra{\overline{\nn_\reI^{(j)}}} \sigma_x \Delta \widehat{U}(x) \ket{\nn_\reI^{(l)} } $, so that $r_{\mathrm{i}} = \int dx Q_{2,1}(x) = O(\Delta x)$ and is small for thin gates.

In the case when $q^2 < 0$, denoting $q = \rmi \kappa$ the reflection and
transmission amplitudes are found to be
\begin{equation}\label{eq:transp_att}
\eqalign{
    r & = -\frac{\sinh(\kappa d)\rme^{\rmi(\alpha + \beta)} }{
            \sinh\left[\kappa d - \rmi (\alpha - \beta) \right]}, \cr
    t & =
        -\frac{\rmi \sin\left(\alpha - \beta \right)}{
            \sinh\left[\kappa d  - \rmi (\alpha - \beta) \right]}.}
\end{equation}
The reflection coefficient monotonously increases to $1$ with the thickness
of the barrier, while the transmission decreases asymptotically
exponentially to zero.

The transition between forms \eref{eq:tbe_form1} and \eref{eq:tbe_form2}
occurs
through the point $q=0$, where $\nn_\reII^{(1)} = \nn_\reII^{(2)}$ and, as a
result, $\det(\widehat{T}_{\reII,\reI})=0$. As has been discussed
above, the reason of the singularity is that the plane wave representation of
solutions of \eref{eq:start_eq} doesn't exhaust all of them. In order to
recover the missing state and to derive the correct form for the transfer
matrix we need to analyze closer Eq.~\eref{eq:start_eq} in the case when
$\widetilde{\epsilon}^2 = h_y^2 + h_z^2$.

Equation~\eref{eq:start_eq} can be rewritten as
\begin{equation}\label{eq:start_proj}
  \sigma_x p_x \psi = 2 \widetilde{\epsilon} \widehat{P} \psi,
\end{equation}
where $\widehat{P} = (\widetilde{\epsilon} - \bosi \cdot
\bi{h}_{yz})/(2\widetilde{\epsilon})$ and $\bi{h}_{yz} = (0,h_{\reII;y},
h_{\reII;z})$.

We notice that $\det(2 \widetilde{\epsilon} \widehat{P}) = q^2$ and,
moreover, when $q=0$, one has $\widehat{P}^2 = \widehat{P}$, thus
$\widehat{P}$ is a projector. The components of $\bi{h}_{yz}$ are real and
therefore the eigenstates of $\widehat{P}$ are $\ket{\pm\nn_{yz}}$, where
$\nn_{yz} = \bi{h}_{yz}/\widetilde{\epsilon}$ and therefore $\widehat{P} =
\ket{-\nn_{yz}}\bra{-\nn_{yz}}$. Taking into account that $\widehat{P}$ can
be diagonalized by rotating the coordinate system around the $x$-axis, we
obtain the general solution of Eq.~\eref{eq:start_proj} for the case $q=0$:
\begin{equation}\label{eq:sol_sec}
  \ket{\psi} = \ket{\psi_0} - 2\widetilde{\epsilon} x \ket{\nn_{yz}}
  \bra{-\nn_{yz}} \psi_0 \rangle,
\end{equation}
where $\ket{\psi_0}$ is an arbitrary spin state. The second terms in this
expression is of secular form and is missed in the representation in terms of
plane waves.

Enforcing the continuity at the boundaries of the barrier we find that the
transfer matrix through the barrier in the case $q=0$ has the form
\begin{equation}\label{eq:ttot_q0}
  \widehat{T}_{\mathrm{tot}} = \widehat{1} - 2 \widetilde{\epsilon} d
        \widehat{M}^{-1}
        \left(
          \begin{array}{cc}
            0 & 0 \\
            \bra{ - \nn_{yz}} \nn_{\reI}^{(1)} \rangle & \bra{ - \nn_{yz}}
            \nn_{\reI}^{(2)} \rangle \\
          \end{array}
        \right),
\end{equation}
where
\begin{equation}\label{eq:Mmat}
  \widehat{M} =
  \left(
          \begin{array}{cc}
            \bra{ - \nn_{yz}} \nn_{\reI}^{(1)} \rangle & \bra{ - \nn_{yz}}
            \nn_{\reI}^{(2)} \rangle \\
            \bra{\nn_{yz}} \nn_{\reI}^{(1)} \rangle & \bra{\nn_{yz}}
            \nn_{\reI}^{(2)} \rangle \\
          \end{array}
        \right).
\end{equation}

Employing the symmetry of involved vectors we obtain
\begin{equation}\label{eq:ttotsec}
  \widehat{T}_{\mathrm{tot}} = \widehat{1} + \rmi \frac{d}{d_c}
  \left(
    \begin{array}{cc}
      -1 & -\rme^{-\rmi \varphi} \\
      \rme^{-\rmi \varphi} & 1 \\
    \end{array}
  \right),
\end{equation}
where $\varphi = A(-\nn_{yz}, \nn_{\reI}^{(1)})$ and
\begin{equation}\label{eq:sec_dc}
  d_c = \frac{1}{\widetilde{\epsilon}}
    \sqrt{\frac{1 + \nn_{yz} \cdot \nn_{\reI}^{(1)}}{1 - \nn_{yz} \cdot
    \nn_{\reI}^{(1)}}} =
  \frac{1}{\widetilde{\epsilon}} \tan(\gamma/2),
\end{equation}
where $\gamma$ is the angle between $\nn_{yz}$ and $\nn_{\reI}^{(1)}$. Using
Eq.~\eref{eq:ttotsec} we find
\begin{equation}\label{eq:rt_sec}
\eqalign{
    r = - \frac{\rme^{-\rmi \varphi}}{1 - \rmi d_c/d}, \cr
    t = \frac{1}{1+ \rmi d/d_c}.
}
\end{equation}

Thus the transition from over-barrier regime ($q^2 > 0$) to canonical tunneling,
characterized by exponential decay with the width of the barrier ($q^2 < 0$), occurs
through the Lorentzian decay with the characteristic length scale $d_c$.

\section{Matrix potential barriers as waveguides}
\label{sec:waveguiding}

We apply the developed technique to analysis of waveguiding properties of
the barrier, or, equivalently, of states localized on the barrier. In general the barrier supports two kinds of such states differing by the
structure of the fermion state inside the barrier. These are either
propagating states, which we will call bulk states, so that $k_{\reII;
x}^{(1,2)} = \pm q$ with $q^2
>0$, or edge states with $k_{\reII; x}^{(1,2)} = \pm \rmi \kappa_{\reII}$ and
$\kappa_{\reII}^2 >0$. 

All localized states are characterized by exponential decay of the wave
function away from the barrier with the rate $\kappa_{\reI} = \sqrt{k_y^2 -
\epsilon^2}$. This implies that at $x< x_\mathrm{L}$ the fermion state is
given by $\ket{\nn_{\reI}^{(2)}}$, while at $x > x_\mathrm{R}$ the state is
$\ket{\nn_{\reI}^{(1)}}$. In order to support such localized state the
transfer matrix through the barrier should satisfy
\begin{equation}\label{eq:loc_cond_gen}
  \bra{\overline{\nn_{\reI}^{(2)}}} \widehat{T}_{\mathrm{tot}}
    \ket{\nn_{\reI}^{(2)}} = 0.
\end{equation}
Using Eqs.~\eref{eq:inside_tm} and \eref{eq:tbe_form} this condition cant be
written as
\begin{equation}\label{eq:loc_cond_overlap}
  \rme^{2\rmi qd} = \frac{\bra{\overline{\nn_{\reII}^{(1)}}} \left.
  \nn_{\reI}^{(2)}\right\rangle
                    \bra{\overline{\nn_{\reII}^{(2)}}} \left.
                    \nn_{\reI}^{(1)}\right\rangle }{
                    \bra{\overline{\nn_{\reII}^{(1)}}} \left.
                    \nn_{\reI}^{(1)}\right\rangle
                    \bra{\overline{\nn_{\reII}^{(2)}}} \left.
                    \nn_{\reI}^{(2)}\right\rangle}.
\end{equation}
This expression is valid in both cases, $q^2>0$ and $q^2 <0$. When $q^2>0$ it
suggests an interesting interpretation: the phase variation inside the barrier
should match the geometric phase spanned by the spin states inside and outside:
$\gamma_{\mathrm{B}} - \gamma_{\mathrm{G}} = \pi m$ with integer $m$, where
$\gamma_{\mathrm{B}} = q d$ and $\gamma_{\mathrm{G}} = A(\nn_{\reII}^{(1)}, -
\nn_{\reI}^{(1)}, -\nn_{\reI}^{(2)})$. As we will show such interpretation in
some generalized form is valid also in the case of edge states.

In order to present Eq.~\eref{eq:loc_cond_overlap} in terms of the parameters
of the system it is more convenient to use an alternative representation of the
transfer matrix using dilation operators.

First, we consider the case $q^2 > 0$. The diagonal form of the transfer matrix
inside the regions with the constant potential implies the ``spectral"
representation
\begin{equation}\label{eq:Ttot_spectral}
  \widehat{T}_{\mathrm{tot}} = \ket{\nn_{\reII}^{(1)}}
  \bra{\overline{\nn_{\reII}^{(1)}}} \rme^{\rmi qd}
                + \ket{\nn_{\reII}^{(2)}} \bra{\overline{\nn_{\reII}^{(2)}}}
                \rme^{-\rmi qd}
\end{equation}
with the matrix elements $\left(\widehat{T}_{\mathrm{tot}}\right)_{i,j} =
\bra{\overline{\nn_{\reI}^{(i)}}} \widehat{T}_{\mathrm{tot}}
\ket{\nn_{\reI}^{(j)}}$. Taking into account Eq.~\eref{eq:n_dil_prop} and the
definition of the dual basis, $\widehat{T}_{\mathrm{tot}}$ can be presented as
\begin{equation}\label{eq:Ttot_dill_prop}
  \widehat{T}_{\mathrm{tot}} = \rme^{b_{\reII} \bi{h}_{yz} \cdot \bosi} \rme^{\rmi q d
  \sigma_x \sign(\widetilde{\epsilon}) }
            \rme^{-b_{\reII} \bi{h}_{yz} \cdot \bosi},
\end{equation}
where $\cosh(b_{\reII} h_{yz}) = |\widetilde{\epsilon}|/q$ and $\sinh(b_{\reII} h_{yz}) = h_{yz} \sign(\widetilde{\epsilon})/q$. In particular,
the zero of reflectivity corresponds to $\bra{\overline{\nn_{\reI}^{(2)}}}
\widehat{T}_{\mathrm{tot}} \ket{\nn_{\reI}^{(1)}} = 0$ and, hence,
\begin{equation}\label{eq:Ttot_dil_zero_r}
 \bra{-\bi{e}_x} \rme^{-b_{\reI} \sigma_y/2} \widehat{T}_{\mathrm{tot}}
 \rme^{b_{\reI} \sigma_y/2} \ket{\bi{e}_x} = 0,
\end{equation}
where $\cosh(b_{\reI}) = \epsilon/k_{\reI;x}^{(1)}$ and $\sinh(b_{\reI}) = k_y/k_{\reI;x}^{(1)}$. It can be seen that Eq.~\eref{eq:Ttot_dil_zero_r} holds
when the width of the barrier satisfies $qd = \pi n$ with integer $n$ or for an
arbitrary width of the barrier when $\bi{h}_{yz} \cdot \bi{e}_z = 0$ and at the
same time $b_{\reI} = b_{\reII} \bi{h}_{yz} \cdot \bi{e}_y$. These conditions
are, of course, identical to those discussed above.

For the case $q = \rmi \kappa_{\reII}$ we have
\begin{equation}\label{eq:Ttot_dill_att}
  \widehat{T}_{\mathrm{tot}} = \rme^{b_{\reII} \bi{h}_{yz} \cdot \bosi /2}
  \rme^{-\kappa_{\reII} d  \bi{l} \cdot \bosi } 
            \rme^{-b_{\reII} \bi{h}_{yz} \cdot \bosi /2},
\end{equation}
where $\bi{l} = \bi{h}_{yz} \times \bi{e}_x/h_{yz}$, $\sinh(b_{\reII} h_{yz}) = \widetilde{\epsilon}/\kappa_{\reII}$, $\cosh(b_{\reII} h_{yz}) = h_{yz}/\kappa_{\reII}$.

The convenience of representations \eref{eq:Ttot_dill_prop} and
\eref{eq:Ttot_dill_att} is that in both cases $q^2 > 0$ and $q^2 < 0$ the
transfer matrix takes the same form
\begin{equation}\label{eq:Ttot_dill_uni}
  \widehat{T}_{\mathrm{tot}} = \rme^{\rmi d \widetilde{\epsilon} \sigma_x  - d h_{yz}
  \bi{l} \cdot \bosi},
\end{equation}
which allows one to treat bulk and edge states on the same footing.

In order to support the localized state the transfer matrix through the barrier
should correspond to rotating vector $\nn_{\reI}^{(2)}$ so that it is directed
along $\nn_{\reI}^{(1)}$. In the case $q^2>0$ according to
Eq.~\eref{eq:Ttot_dill_prop} this is eventually achieved by the conventional
rotation around the $x$-axis, which yields the geometric interpretation
mentioned above. In the case of edge states the corresponding rotation is
hyperbolic as is illustrated by Eqs.~\eref{eq:n_dil_prop} and
\eref{eq:n_dil_att}. Thus the waveguiding modes supported by the bulk states may occupy multiple bands, while the edge states may support only the single band.

First, we analyze Eq.~\eref{eq:loc_cond_gen} for the case of edge states,
i.e. when $\kappa_{\reII} > 0$. Presenting Eq.~\eref{eq:loc_cond_gen} as
\begin{equation}\label{eq:loc_cond_dill}
  \bra{\bi{e}_z} \rme^{- b_{\reI} \sigma_y/2} \widehat{T}_{\mathrm{tot}}
    \rme^{b_{\reI} \sigma_y/2} \ket{\bi{e}_z} = 0,
\end{equation}
where $\sinh(b_{\reI}) = \epsilon/\kappa_{\reI}$ and $\cosh(b_{\reI}) = k_y/\kappa_{\reI}$, and expanding $\widehat{T}_{\mathrm{tot}}$ we obtain
the condition of localization in the form
\begin{equation}\label{eq:loc_cond_exp_att}
  \gamma_{\mathrm{G}} = \gamma_{\mathrm{B}},
\end{equation}
where $\gamma_{\mathrm{B}} = \kappa_{\reII} d$, and
\begin{equation}\label{eq:loc_cond_exp_att_tanh}
  \tanh(\gamma_{\mathrm{G}}) = \frac{\kappa_{\reI}\kappa_{\reII}}{D}
\end{equation}
with $D= \epsilon\widetilde{\epsilon} - k_y (k_y + U_y)$.

\begin{figure}
  \centering
  \includegraphics[width=3in]{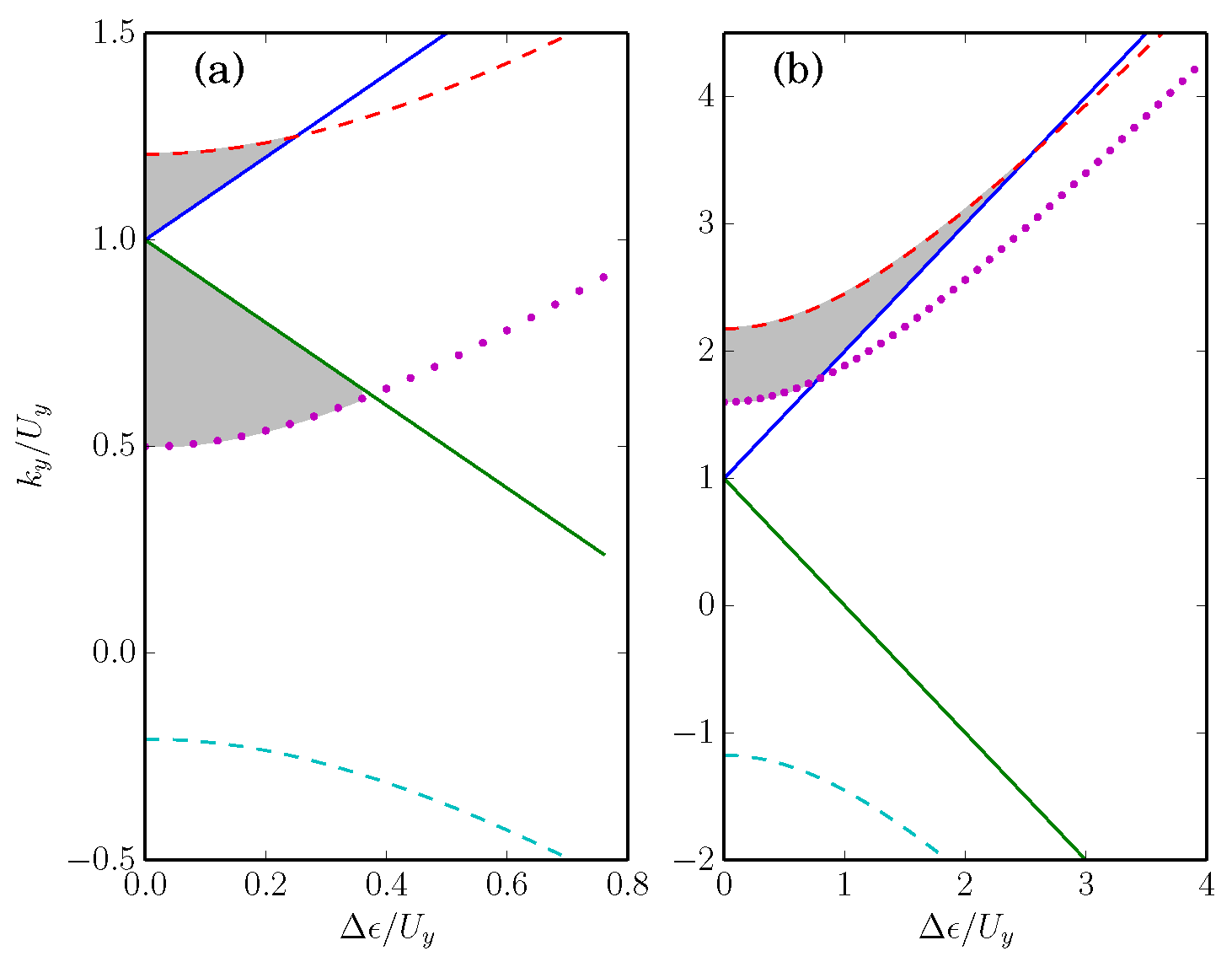}
  \caption{Characteristic curves on $(k_y, \Delta \epsilon)$-plane, where
  $\Delta \epsilon = \sqrt{\widetilde{\epsilon}^2 - U_z^2}$, determining the structure of the spectrum of
  edge states: $k^{(\pm)}(\Delta\epsilon)$ or $\kappa_{\reII}(k_y, \Delta\epsilon) = 0$ (solid line),
  $D(k_y, \Delta\epsilon) = 0$ (dashed line) and $\kappa_{\reI}(k_y, \Delta\epsilon) = 0$ (dotted line) plotted
  for the case $U_y < 0$ and intermediate $V$.
  The edge states may exist only in shaded areas.
  In the case $V+ |U_z| < |U_y|$ (a), the edge state can be either in the upper region,
  $k_y > |U_y|$, or the lower region depending on the relation between $d$
  and $d_c$ (see the main text). When $V + |U_z| > |U_y|$ (b)
  the band edge is determined by $\left.
\gamma_{\mathrm{B}}'/\gamma_{\mathrm{G}}'\right|_{k_y = k^{(+)}} >
1$.}\label{fig:ints_edge}
\end{figure}

The general structure of the spectrum of edge states is determined by the overlap of intervals, where
$k_y$ may reside in order to have positive $\kappa_{\reI}$, $\kappa_{\reII}$
and $D$, as is illustrated in figure~\ref{fig:ints_edge}. The general form of the spectrum is
determined by simple relations between $V$ and $U_y$. It can be seen that
solutions of Eq.~\eref{eq:loc_cond_exp_att} with $k_y
>0$ and $k_y < 0$ may exist only when $V < -U_y$ and $V < U_y$,
respectively. Thus, for sufficiently deep attracting barriers, $V<0$ and $|V| >
|U_y|$, Eq.~\eref{eq:loc_cond_exp_att} may support for the same energy
solutions with both $k_y > 0$ and $k_y < 0$. When $V$ increases so that $|V| <
|U_y|$, for a particular energy there may be only one solution and $k_y$ and
$U_y$ must be of opposite signs. With further increase of $V$, in sufficiently strong repulsive potentials 
$V > |U_y|$ no solutions of Eq.~\eref{eq:loc_cond_exp_att} exist. 

We would like to emphasize that the condition of existence of waveguiding modes supported by the edge states, $V < |U_y|$, does not require the scalar potential to be attractive nor the presence of the mass gap (i.e. when $U_z \ne 0$).

In order to avoid overly cumbersome expressions we discuss details of the
spectrum in the case when $V = 0$ and we take for definiteness $U_y < 0$, so
that only solutions with $k_y > 0$ may exist. As can be seen from
figure~\ref{fig:ints_edge} the maximal energy of the edge states cannot exceed
the value determined by the intersection of the curve $\kappa_{\reII}(k_y,
\epsilon) = 0$ with either $D(k_y, \epsilon) = 0$ (if $|U_y/U_z| < \sqrt{3}$)
or with $\kappa_{\reI}(k_y, \epsilon) = 0$ (if $|U_y/U_z| > \sqrt{3}$):
\begin{equation}\label{eq:max_energy_edge}
  \epsilon < \epsilon_{\mathrm{max}} =
    \max\left((U_y^2 - U_z^2)/2, U_z\sqrt{U_y^2 + U_z^2}\right)/U_y.
\end{equation}

In order to find the actual width of the band of edge states we consider the
condition of existence of solutions of Eq.~\eref{eq:loc_cond_exp_att} within
the intervals shown in figure~\ref{fig:ints_edge}. Depending on whether $d <
d_{\mathrm{c}}$ or $d > d_{\mathrm{c}}$, where $d_{\mathrm{c}} =
\mathrm{Re}\left[\sqrt{U_y^2 - U_z^2}/U_z^2\right]$, the condition has the form
$\left. \gamma_{\mathrm{B}}'(k_y)/\gamma_{\mathrm{G}}'(k_y)\right|_{k_y =
k^{(-)}} < 1$ or $\left.
\gamma_{\mathrm{B}}'(k_y)/\gamma_{\mathrm{G}}'(k_y)\right|_{k_y = k^{(+)}} >
1$, respectively. Here
\begin{equation}\label{eq:k_app_edge}
  k^{(\pm)} = -U_y \pm \sqrt{\widetilde{\epsilon}^2 - U_z^2}
\end{equation}
are zeros of $\kappa_{\reII}(k_y)$ for $|\widetilde{\epsilon}| > |U_z|$
(this expression is valid in the case $V \ne 0$ as well). Thus we find that
the width of the band of edge states is
\begin{equation}\label{eq:width_edge}
  \Delta_{\mathrm{E}} = \sqrt{U_z^2 + \delta_{\mathrm{E}}^2},
\end{equation}
where
\begin{equation}\label{eq:top_edge_wide}
  \delta_{\mathrm{E}} = \frac{U_z^2}{U_y} - \frac{1}{U_y
  d^2}\left(\sqrt{1+d^2(U_y^2+U_z^2)} -1 \right).
\end{equation}

The characteristic form of $k_y(\epsilon)$ is linear in the low energy limit
  $v_{\mathrm{E}} k_y = \epsilon$,
where
\begin{equation}\label{eq:edge_low_speed}
  v_{\mathrm{E}}^{-1} = 1 +
        \frac{U_y^2}{2(U_y^2 + U_z^2)} \tanh^2\left(d\sqrt{U_y^2 +
        U_z^2}\right).
\end{equation}
Thus the edge states are massless excitations.

Now we turn to the analysis of bulk states, i.e. states localized inside the
barrier and characterized by $q^2 >0$, whose spectrum has much richer
structure. In this case condition of localization~\eref{eq:loc_cond_dill}
takes the form
\begin{equation}\label{eq:loc_cond_prop}
  \gamma_{\mathrm{B}} - \gamma_{\mathrm{G}} = \pi n,
\end{equation}
where $\gamma_{\mathrm{B}} = q d$ and
\begin{equation}\label{eq:tan_G}
  \tan(\gamma_{\mathrm{G}}) = \frac{\kappa_{\reI} q}{\epsilon
  \widetilde{\epsilon} - k_y(k_y + U_y)}.
\end{equation}
In contrast to the previous case, the phases should match up to multiples of
$\pi$. According to Eq.~\eref{eq:Ttot_dill_prop}, this corresponds to
different number of full rotations of the incoming spin state inside the
barrier. States corresponding to $n=0$
constitute the fundamental band and those with $n >0$ form higher bands.

\begin{figure}
  \centering
  \includegraphics[width=3in]{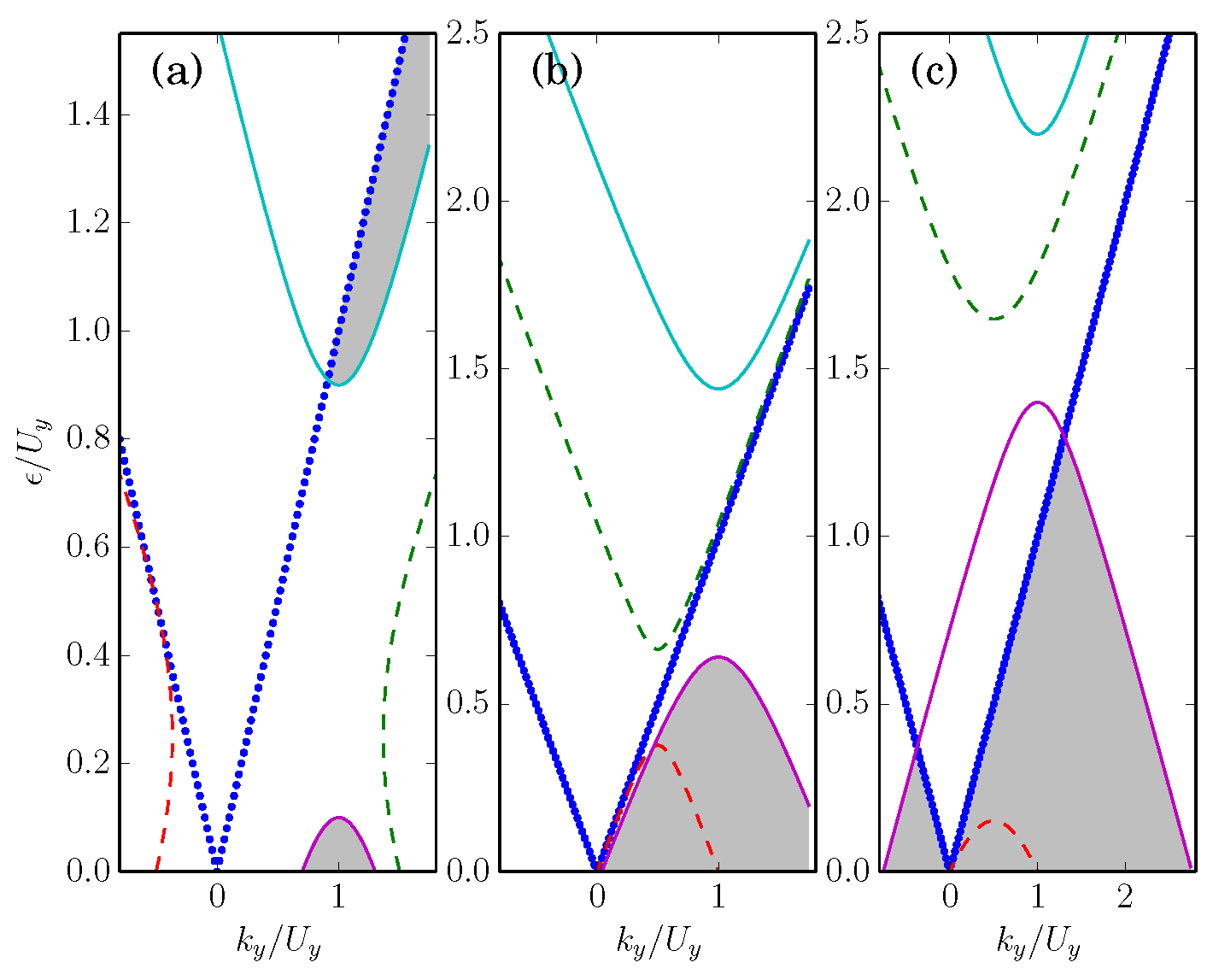}\\
  \caption{The structure of the spectrum of localized states propagating across
  the barrier (bulk modes). In all three panels the localized states may exist
  only within the shaded regions, dotted lines correspond to
  $\kappa_{\reI}(\epsilon, k_y) = 0$, i.e. $\epsilon = |k_y|$, the solid lines show
  $q(\epsilon, k_y) = 0$, i.e. $\epsilon = V \pm \sqrt{(k_y + U_y)^2 +U_z^2}$ for $U_z < 0$.
  The dashed lines show $D(\epsilon, k_y) = 0$, where $\gamma_{\mathrm{G}} = \pi/2$.
  Its position allows one to estimate the variation of the geometric phase along
  the line connecting the opposite sides of a shaded region at fixed energy and
  to formulate the condition of existence of a solution of
  Eq.~\eref{eq:loc_cond_prop}. (a) $V < |U_y|$. The case $V > |U_z|$ is shown
  (more specifically, $|U_z/U_y| = 0.4$ and $V/|U_y| = 0.4$), when the finite
  band may exist in sufficiently wide barriers. (b) When $V < |U_y|$
  the infinite band disappears, while the finite band may exist if $V < |U_z|$.
  The case $|U_z/U_y| = 0.4$ and $V/|U_y| = 1.04$ is shown.
  (c) When $V > \sqrt{U_y^2 + U_z^2}$ there may be finite bands corresponding
  to $k_y$ of both signs. The barrier looses the property of uni-directionality.
  The case $|U_z/U_y| = 0.4$ and $V/|U_y| = 1.8$ is shown.
   }\label{fig:bulk_regimes}
\end{figure}

The dependence of the spectrum of localized states on the relation between
$V$ and $U_y$ is more complex, than in the previous case. Let us assume for
concreteness that $U_y < 0$. An analysis of conditions $q^2 > 0$ and
$\kappa^2_{\reI} > 0$ (see figure~\ref{fig:bulk_regimes}) shows that there
are three possibilities:
\begin{enumerate}
\item $V < |U_y|$. There are no solutions with $k_y < 0$, while states with
    $k_y > 0$ occupy an infinite band possibly with a gap (see
    figure~\ref{fig:bulk_regimes}(a).
\item $|U_y| < V$. The infinite band disappears. A finite band of states with
    $k_y> 0$ may exist if additionally $V > |U_z|$ (see
    figure~\ref{fig:bulk_regimes}(b)). Thus, if $|U_y| < V < |U_z|$, there are
    no localized states, either bulk or edge.
\item $\sqrt{U_y^2 + U_z^2} < V$. If there are solutions, they exist for both
    $k_y > 0$ and $k_y < 0$ occupying bands of finite size (see
    figure~\ref{fig:bulk_regimes}(c)).
\end{enumerate}

Whether there exists a solution of Eq.~\eref{eq:loc_cond_prop} at
a chosen energy depends on details of the variation of
$\gamma_{\mathrm{B}}(k_y)$ and $\gamma_{\mathrm{G}}(k_y)$. The latter, in
turn, depends on the position of the pole of Eq.~\eref{eq:tan_G}, depicted
by dotted lines in figure~\ref{fig:bulk_regimes}. This leads to a cumbersome
system of conditions and, therefore, we limit ourselves to the case $U_z =
0$ noticing that the main effect of $U_z \ne 0$ is separating regions, where
$q^2 > 0$ as is demonstrated by figure~\ref{fig:bulk_regimes}(a).

In the case $V< U_y$ the existence of solutions is determined by the
condition $\left.
\gamma_{\mathrm{B}}'(k_y)/\gamma_{\mathrm{G}}'(k_y)\right|_{k_y = k^{(-)}} >
1$, where $k^{(-)}$ is given by Eq.~\eref{eq:k_app_edge}. This condition is
satisfied, when
\begin{equation}\label{eq:bulk_up}
  \epsilon > \epsilon_{\mathrm{U}} = V + \frac{V - |U_y|}{
            1 + \sqrt{1 + d^2 (U_y^2 - V^2)}}
\end{equation}
and
\begin{equation}\label{eq:bulk_down}
  0 < \epsilon < \epsilon_{\mathrm{D}} = V - \frac{V + |U_y|}{
            1 + \sqrt{1 + d^2 (U_y^2 - V^2)}}.
\end{equation}
These inequalities define two bands formed by bulk states. One band extends
to infinity, while another, existing when $V > 1/d$, is finite.

The infinite band consists of overlapping fundamental band and higher bands.
The dispersion law of the fundamental near $\epsilon_{\mathrm{U}}$ is
linear, thus the respective excitations are massless. The spectrum of higher
bands, however, shows an interesting feature. Let us consider the $n$-th
band with $n>1$. The form of solutions of Eq.~\eref{eq:loc_cond_prop}
essentially depends on whether $\epsilon > \epsilon_{\mathrm{M}}$ or
$\epsilon < \epsilon_{\mathrm{M}}$, where $\epsilon_{\mathrm{M}} = (|U_y|
+V)/2$ is the minimal energy such that $\epsilon > k^{(-)}$. In the general
case the region of massive bands is given by $\epsilon_{\mathrm{M}} = (U_y^2
+ U_z^2 - V^2)/2 (|U_y| - V)$.  If $\epsilon < \epsilon_{\mathrm{M}}$, the
position of the bottom of the $n$-th band, $\epsilon_{\mathrm{U}}^{(n)}$,
can be estimated as
\begin{equation}\label{eq:n_band_up}
\epsilon_{\mathrm{U}}^{(n)} \approx V + \Delta^{(n)},
\end{equation}
where $\Delta^{(n)} = {\pi n}/{d}$. The dispersion law of the $n$-th band
near $\epsilon_{\mathrm{U}}^{(n)}$ has the form
\begin{equation}\label{eq:disp_n_up}
  \frac{(k_y - |U_y|)^2}{2 \mu^{(n)}} = \epsilon -
  \epsilon_{\mathrm{U}}^{(n)},
\end{equation}
where $\mu^{(n)} = \pi (m+1/2)/|U_y| d$. Thus, only higher bands with
numbers $n \leq n_{\mathrm{M}} = \left[(|U_y| - V)d/2\pi\right]$, where
$\left[\ldots\right]$ denotes taking the integer part, are massive.

The finite band occupying $0 < \epsilon < \epsilon_{\mathrm{D}}$ is also
formed by overlapping fundamental band with a finite number (possibly zero)
of higher bands. The position of the top of the $n$-th band is
$\epsilon_{\mathrm{D}}^{(n)} = V - \Delta^{(n)}$. Thus the number of higher
bands contained in the low-energy finite band is $N = \left[d V /\pi
\right]$.

The interesting feature of the finite band is that excitations near the top
of all higher bands bands are massive and their masses match masses of the
respective excitations in the infinite band but are negative. The dispersion
laws are of the form $(k_y - |U_y|)^2/2\mu^{(n)} =
\epsilon_{\mathrm{D}}^{(n)} - \epsilon$.

Only when $V > \sqrt{U_y^2 + U_z^2}$, the barrier may admit localized bulk
states with $k_y$ of the same sign as $U_y$. The bands occupied by states
with $k_y < 0$ and $k_y < 0$ are, however, of different size, while contain
approximately the same number of bands. For a given energy $\epsilon$ states
with positive and negative $k_y$ are inside intervals $(\epsilon, k^{(+)})$
and $(k^{(-)}, -\epsilon)$, respectively, where $k^{(\pm)}$ are given by
Eq.~\eref{eq:k_app_edge}. In order to estimate the position of the $n$-th
band, with $n=0, 1, \ldots$, we approximate $\gamma_{\mathrm{G}} \approx
\pi$, which is a good approximation when $V$ significantly exceeds
$\sqrt{U_y^2 + U_z^2}$ and $\epsilon$ is not too close to zero. Thus for
$k_y < 0$ we find
\begin{equation}\label{eq:neg_strong}
  \epsilon_-^{(n)} = \frac{1}{2} \left(V - |U_y|\right) -
        \left(\frac{\pi(n+1)}{d}\right)^2 \frac{1}{2(V + |U_y|)}.
\end{equation}
Within the adopted approximation for $\gamma_{\mathrm{G}}$ the dispersion
laws of the bands are approximately linear, implying massless excitations.
This approximation, however, breaks in the immediate vicinity of the top
points of the bands.

The same approximation can be used for studying states with positive $k_y$
yielding for the top of the $n$-th band
\begin{equation}\label{eq:pos_strong}
  \epsilon_+^{(n)} = \frac{1}{2} \left(V + |U_y|\right) -
        \left(\frac{\pi(n+1)}{d}\right)^2 \frac{1}{2(V - |U_y|)}.
\end{equation}

It should be noted that Eqs.~\eref{eq:neg_strong} and \eref{eq:pos_strong}
predict the same number of bands with positive and negative $k_y$. While
this result is obtained using a crude approximation $\gamma_{\mathrm{G}}
\approx \pi$, it breaks only in barriers with carefully chosen parameters,
in which states with negative $k_y$ may have one band less than the states
with $k_y > 0$. Equations~\eref{eq:neg_strong} and \eref{eq:pos_strong} also
correctly predict that not all barriers with strong $V$ may support
localized states. More accurate estimate for parameters of the barrier
allowing at least fundamental band can be obtained as
$\gamma_{\mathrm{B}}(\epsilon = 0, k_y = 0) > \pi/2$ yielding  $V^2 - U_y^2
- U_z^2 > (\pi/2d)^2$.

\section{Application for topological insulators}
\label{sec:topological}

The results of the previous sections can be directly applied for description of
electrons moving along the surface of a topological insulator. In order to do
this, two important circumstances should be taken into account. First, the
Hamiltonian of a free electron in this case is usually taken in the Rashba form
  $H_{\mathrm{R}} = v \bi{e}_z \cdot (\bosi \times \bi{p})$,
which is different from the electron Hamiltonian in graphene: $H_{\mathrm{W}} =
v \bosi \cdot \bi{p}$. Hamiltonians $H_{\mathrm{R}}$ and $H_{\mathrm{W}}$,
however, are equivalent up to different choices of the $\sigma$-matrices,
generators of the $su(2)$ algebra. Second, a consistent treatment of the matrix
potential requires taking into account the following circumstance. If, for instance, the
matrix potential is implemented by a vector potential, there's a strong
magnetic field at the points of strong variation of the vector potential. The
Zeeman interaction of the electron spin with this magnetic field  cannot be
neglected and has to be taken into account.

In order to keep the general character of the consideration and to distinguish the effect of the Zeeman interaction, we formally distinguish the contribution of the matrix and vector potentials and, thus, consider the equation of motion of the form
\begin{equation}\label{eq:TI_eq_mot}
  \left[\bi{e}_z \cdot (\bosi \times (\bi{p} - e \bi{A})) + \widehat{U}\right]
  \psi = \epsilon \psi,
\end{equation}
where $\bi{A}$ is the vector potential. The matrix potential $\widehat{U}$
can be presented in terms of $\sigma_{x, y, z}$
\begin{equation}\label{eq:matr_pot_TI}
  \widehat{U} = V \widehat{1} + {\bi{U}'} \cdot \bosi
\end{equation}
with ${\bi{U}'} = \bi{U} + g\bi{B}$, $\bi{B} = \nabla \times \bi{A}$
is the magnetic field and $g$ is the gyromagnetic ratio. In order to show the
equivalence of
Eq.~\eref{eq:TI_eq_mot} and \eref{eq:start_eq} we introduce
$\widetilde{\sigma}_x = -\sigma_y$ and $\widetilde{\sigma}_y = \sigma_x$,
which correspond to the representation $\widetilde{\sigma}_x \bi{e}_x +
\widetilde{\sigma}_y \bi{e}_y = \bi{e}_z \times \left({\sigma}_x \bi{e}_x +
{\sigma}_y \bi{e}_y\right)$. It can be easily checked that
$\widetilde{\sigma}_x$, $\widetilde{\sigma}_y$ and $\sigma_z$ satisfy the
same commutation relations as ${\sigma}_x$, ${\sigma}_y$ and $\sigma_z$. In
terms of $\widetilde{\bosi} = (\widetilde{\sigma}_x, \widetilde{\sigma}_y,
\sigma_z)$ Eq.~\eref{eq:TI_eq_mot} is written as
\begin{equation}\label{eq:TI_eq_rotated}
  \widetilde{\bosi} \cdot \left(\bi{p} + \widetilde{\bi{U}} \right) \psi =
  (\epsilon - V)\psi,
\end{equation}
where $\widetilde{\bi{U}} = - e \bi{A}+ \widetilde{\bi{U}}_\perp + U_z \bi{e}_z$
and $\widetilde{\bi{U}}_\perp = \bi{e}_z \times \bi{U}'$.
Equation~\eref{eq:TI_eq_rotated} has the same form as Eq.~\eref{eq:start_eq}
but with vectors and spin states rotated by $\pi/2$ around the $z$-axis. Thus
the consideration of Section~\ref{sec:e_states} can be simply repeated in the
present case. In order to restore the directions of the effective fields and
spin states for the electron in topological insulator one only needs to perform
the inverse rotation, i.e. rotate the respective vectors by $-\pi/2$ around the
$z$-axis. Having this relation established we will ommit tilde while writing the components of the effective matrix potential.

In order to obtain the transfer matrix, however, it is necessary to account for
the effect of the Zeeman term. We introduce $f(x) = \theta(x - x_L)\theta(x_R -
x)$, where $\theta(x)$ is the Heaviside step function, and denote $\bi{A}(x) =
\bi{A} f(x)$, so that $\bi{B}(x) = \nabla f \times \bi{A}$. In the immediate
vicinity of the boundary of the barrier one can neglect non-singular
contributions in Eq.~\eref{eq:TI_eq_rotated} thus obtaining
\begin{equation}\label{eq:TI_bound}
  - \rmi \widetilde{\sigma}_x \frac{\partial \psi}{\partial x}  +
        g \frac{\partial f}{\partial x} \left(\sigma_z A_y + \widetilde{\sigma}_x A_z\right) \psi = 0.
\end{equation}
The solution of this equation can be written as
\begin{equation}\label{eq:TI_sol_bound}
  \psi(x_2) = \exp \left\{ g (\widetilde{\sigma}_y A_y - i A_z)
  [f(x_2) - f(x_1)]\right\} \psi(x_1).
\end{equation}
It is seen that at $x = x_L$ and $x_R$ the spin experiences discontinuity
described by the dilation operators $\rme^{g \widetilde{\sigma}_y A_y}$ and $\rme^{-g
\widetilde{\sigma}_y A_y}$, respectively. These jumps are conveniently
accounted for in the representation of the transfer matrix through the barrier,
$\widehat{T}_{\mathrm{tot}}$, in terms of dilation operators:
\begin{equation}\label{eq:TI_T_tot}
  \widetilde{T}_{\mathrm{tot}} =
        \rme^{-g \widetilde{\sigma}_y A_y}
            \rme^{i d \widetilde{\epsilon} \widetilde{\sigma}_x - d h_{yz} \bi{l} \cdot \bosi}
        \rme^{g \widetilde{\sigma}_y A_y}.
\end{equation}
Comparing with expressions for matrix elements of $\widehat{T}_{\mathrm{tot}}$,
see e.g. Eqs.~\eref{eq:Ttot_dil_zero_r} and \eref{eq:loc_cond_dill}, it can be
seen that the effect of the spin jump reduces to a straightforward modification
of the dilation operator determining the incoming spin state: $b_{\reI} \to
\widetilde{b}_{\reI} = {b}_{\reI} + 2g A_y$. This allows us to apply directly the
results of the previous sections.

First we consider the modification of the KT condition. It has the same form as
Eq.~\eref{eq:Ttot_dil_zero_r}, which results in
\begin{equation}\label{eq:TI_KT_cond}
  \sinh(2 g A_y)\left[\epsilon \widetilde{\epsilon} - k_y (k_y + U_y)\right] -
        \cosh(2 g A_y)\left[k_y V + U_y \epsilon\right] = 0.
\end{equation}
This equation determines the dependence of the direction of zero reflectivity
for arbitrary barrier width on parameters of the barrier:
\begin{equation}\label{eq:TI_KT_ky}
\eqalign{
  k_y = & -\frac{1}{2}\left(V \coth(2 g A_y) +U_y\right) \cr
  & \pm \sqrt{(U_y + V \coth(2 g A_y))^2 - 4 \epsilon(U_y \coth(gA_y) - \widetilde{\epsilon})}.
}
\end{equation}
We briefly analyze this result assuming for concreteness that $A_y > 0$.

\begin{figure}
  \centering
  \includegraphics[width=3in]{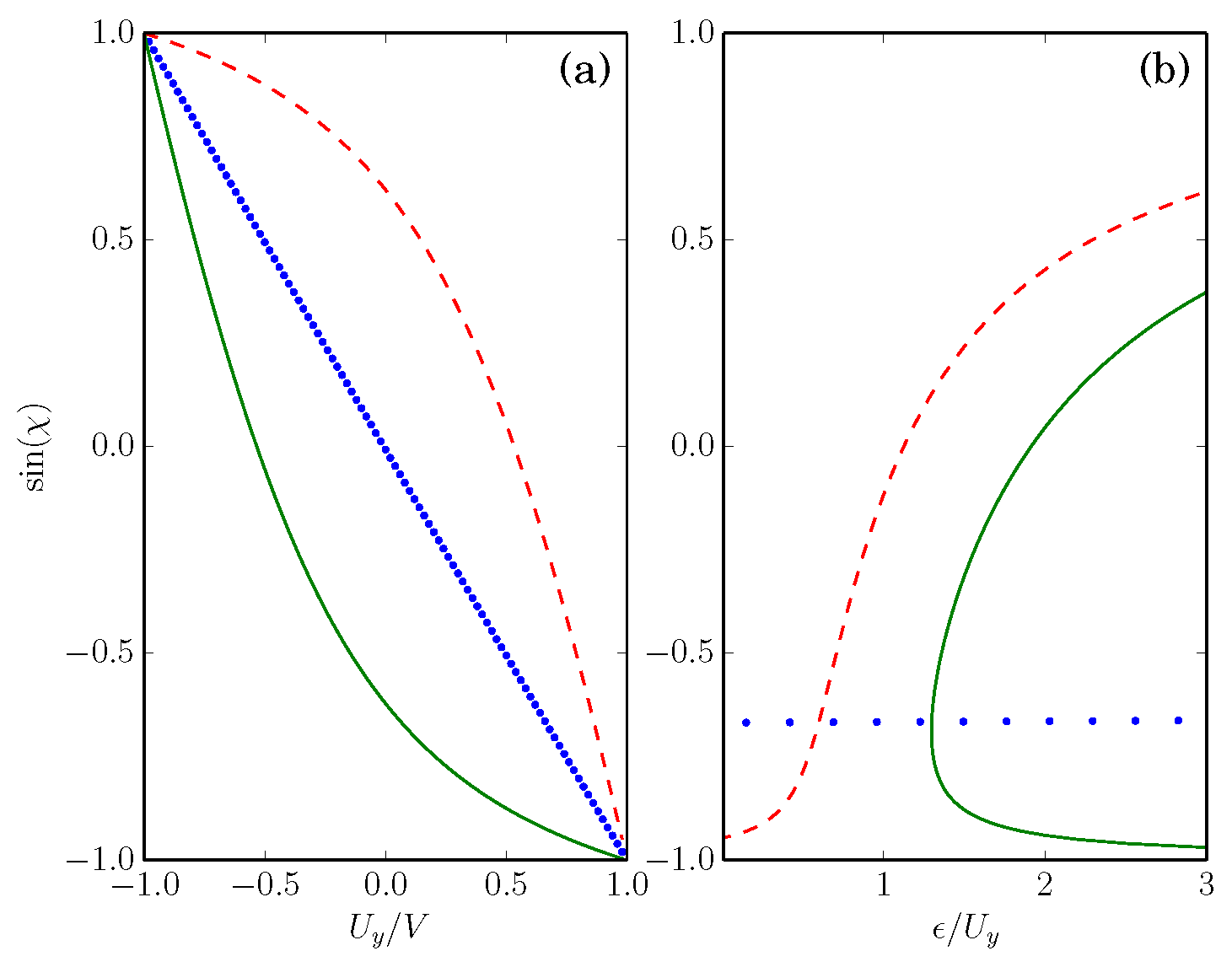}\\
  \caption{The direction corresponding to the Klein tunneling, i.e. zero
  reflectance for an arbitrary width of the barrier. (a) The dependence
  on the relation between $V$ and $U_y$ for $A_y>0$ (solid line), $A_y = 0$
  (dotted line) and $A_y < 0$ (dashed line). (b) The dependence on energy.
  If the spin distribution is continuous, $A_y = 0$, the KT direction
  exists when $V > |U_y|$ and is energy independent (dotted line). In the
  case when $V > |U_y|$ and $A_y \ne 0$ the KT direction demonstrates non-trivial
  dependence on energy (dashed line). If $V < |U_y|$ the KT direction is present
  only when $A_y \ne 0$ (solid line).} \label{fig:TI_KT}
\end{figure}

In the case $\epsilon < V$ the effect of the spin discontinuity is a
modification of the dependence of the KT direction on the relation between
$U_y$ and $V$ as is illustrated in figure~\ref{fig:TI_KT}(a). Additionally the KT
direction becomes energy dependent (see figure~\ref{fig:TI_KT}(b)). At the same
time the condition for the KT to exist remains the same as in the case of
continuous spin distribution, $|U_y| < V$.

When $\epsilon > V$, however, new features appear. First of all, when $|U_y| >
V$ the KT is no longer suppressed but rather appears at high energies,
\begin{equation}\label{eq:KT_energy}
\epsilon > \frac{1}{2}\left(V + U_y\coth(2 g A_y) \right) +
            \frac{1}{2\sinh(2 gA_y)} \sqrt{U_y^2 - V^2}.
\end{equation}
Moreover, if $\epsilon > V/(1-\rme^{-4gA_y})$ there are two distinct KT
directions.

In the similar way the effect of the spin discontinuity on localized modes can
be studied. Using Eq.~\eref{eq:TI_T_tot} in localization
condition~\eref{eq:loc_cond_dill} we find that the spin jump leads to
modification of the geometric phase only
\begin{equation}\label{eq:TI_geom_phase}
  \tanh(\gamma_{\mathrm{G}}) = \frac{\kappa_{\reI} \kappa_{\reII}}{D_{\mathrm{T}}},
\end{equation}
where
\begin{equation}\label{eq:TI_D}
\eqalign{
  D_{\mathrm{T}} = & \widetilde{\epsilon}\left[\epsilon \cosh(2 g A_y) + k_y \sinh(2 g A_y)\right] \cr
        & - (k_y + U_y) \left[k_y \cosh(2 g A_y) +  \epsilon \sinh(2 g A_y)\right].
}
\end{equation}
The equation for localized states $\tanh(\gamma_{\mathrm{B}}) =
\tanh(\gamma_{\mathrm{G}})$ can be analyzed using the same approach as in the
previous section. The effect of the spin discontinuity can be seen to be less
significant than for the KT. The main conditions for existence of localized
states and the boundary between massive and massless modes remain the same as
in the case of continuous spin. The exact positions of the band edges and
masses are modified. The explicit expressions, however, are too cumbersome and
we do not provide them here.

\section{Conclusion}

We present a detailed consideration of the propagation of a 2D Dirac electron in the presence of a rectangular matrix potential barrier. We describe scattering with the help of spinor transfer matrices, which relate the orientation of the electron spin state at different points of the system given in terms of superposition of eigen spin coherent states. We show that the Klein tunneling is suppressed in the presence of the mass term, $\propto \sigma_z$. In the absence of such contribution, the Klein tunneling is not suppressed but is observed at an oblique direction with the angle of incidence determined by the ratio between the scalar and vector components of the matrix potential.

The analysis of scattering is applied for studying of waveguiding
properties of the matrix potential barrier. Depending on the electron energy and parameters of the barrier, it may support states localized near the boundaries (edge states) or penetrating the interior (bulk states). We describe the general properties of the waveguiding modes, determine the widths of the bands and obtain the dispersion laws of the low-energy excitations. We show that both kinds of waveguiding modes, supported by edge and bulk states, may demonstrate the property of uni-directionality when the barrier admits only waveguiding modes with $k_y U_y < 0$. We obtain general conditions governing the general form of the spectrum of waveguiding modes. In particular, we show that in barriers with sufficiently strong attractive scalar potential, the waveguiding modes supported by the bulk states may demonstrate gapped spectrum in wide barriers. When the scalar potential increases, the bottom of the infinite band raises and in sufficiently strong repulsive scalar potential only the finite band remains, which serves as the precursor to loosing the uni-directionality property.

While the waveguiding modes supported by the edge states are massless, those supported by the balk states have both massive and massless bands. We find the energy region, where the massive bands are located, and find their number.

In addition to the case when the electron is characterized by the pseudospin, we consider the case the spin is real and thus is sensitive due to the Zeeman effect to the magnetic field at the boundaries of the barrier. Its the most significant manifestation is the appearance at sufficiently high energies of two distinct directions corresponding to the Klein tunneling.

\ack

The University of Michigan team was supported by the Air Force Office of Scientific Research (AFOSR) Grant No. FA9550-12-1-0402. The University of Central Florida team was supported by the National Science Foundation (NSF) Grants No. ECCS-1128597 and No. ECCS-1514089.

\section*{References}

\bibliographystyle{iopart-num}
\bibliography{weyl_scattering}

\end{document}